# The diffusion dynamics of choice: From durable goods markets to fashion first names

Baptiste Coulmont,[1] Virginie Supervie[2,3] and Romulus Breban[4*]

[1]*Univ Paris 08, CNRS UMR 7217, Paris F75017, France*
[2]*Sorbonne Universités, UPMC Univ Paris 06, UMR_S 1136 Pierre Louis Institute of Epidemiology and Public Health, F-75013, Paris, France*
[3]*INSERM, UMR_S 1136 Pierre Louis Institute of Epidemiology and Public Health, F-75013, Paris, France*
[4]*Institut Pasteur, UEME, Paris 75724, France*
[*]**To whom correspondence should be addressed.** E-mail: Romulus.Breban@pasteur.fr

**Abstract**
Goods, styles, ideologies are adopted by society through various mechanisms. In particular, adoption driven by innovation is extensively studied by marketing economics. Mathematical models are currently used to forecast the sales of innovative goods. Inspired by the theory of diffusion processes developed for marketing economics, we propose, for the first time, a predictive framework for the mechanism of fashion, which we apply to first names. Analyses of French, Dutch and US national databases validate our modelling approach for thousands of first names, covering, on average, more than 50% of the yearly incidence in each database. In these cases, it is thus possible to forecast how popular the first names will become and when they will run out of fashion. Furthermore, we uncover a clear distinction between popularity and fashion: less popular names, typically not included in studies of fashion, may be driven by fashion, as well.

**INTRODUCTION**
Understanding how individuals make choices and how these choices interplay is a fundamental problem, common to many sciences. For more than 40 years, marketing economics has been successfully using mathematical models based on the theory of diffusion processes to predict how individual-level choices to purchase innovative goods spread across society [1-3]. Similarly, we propose a mathematical framework to describe individual-level choices to adopt fashion and apply it to the case of first name data.

Choosing a baby's name appears to be a difficult task: advice from friends and family, countless books, websites, and even name consultants are feeding parent's anxiety to select the right name. These parental concerns are historically recent, as first names were transmitted, rather than chosen, until the beginning of the 19th century [4, 5]. Nowadays, parents explicitly rely on taste-based arguments to explain their choices [5]. Hence, over time, various first names grow into fashion, then peak and go extinct.

Economists, physicists, psychologists and anthropologists have proposed modelling frameworks for the population dynamics of individuals bearing a certain first name. Several key results have thus emerged. The speeds of adoption and abandonment of fashion names appear to be positively correlated [6]. Delay and memory mechanisms have been invoked for explaining these dynamics [7]. Furthermore, analysis of the distribution of first names versus popularity suggested that the process of name choice could be random copying [8-10]. However, the literature on first names has shortcomings:

(a) *Fashion is often taken for popularity*. Although studies (including the historical ones) are often based on nationwide databases [11], they factually focus on the most popular names (top 10, 20 or 100), which are implicitly considered to be fashion names. Evidence that names are subject to fashion is typically drawn from the turnover among top ranked names. The increasing turnover during the last two centuries is interpreted as a societal shift from tradition to fashion. However, popularity is not equivalent to fashion, and focusing on the top ranked names disregards the possibility of fashion at small population scales.

(b) *Quantitative modelling of name dynamics has yet to provide a pragmatic, mechanistic explanation for fashion.* Two major approaches have been used for modelling name dynamics and quantitative fit of first name data. The first approach relies on empirical models using only a few parameters [12]; however, such models have no mechanistic justification. The second approach uses more sophisticated mechanistic models---e.g., for memory and delay [7]---involving numerous parameters, which yet provide limited insight in the sociological mechanisms of name choice and are difficult to use for prediction [7, 13].

Here we propose a new conceptual framework for first name dynamics. We postulate that the spread of many names, called *fashion names*, has similarities to the first purchase of innovative durable goods (e.g., cars, electronics), relying on diffusion processes defined by four key elements: fashion (innovation in case of market goods), communication channels, time and the social system. There are three major arguments in favour of this analogy. First, just like customers purchase durable goods only a few times in their lifetime, parents choose baby names only a few times in their lifetime (on average, two per couple, in a stable population). Second, both names and goods are in competition, subject to individual-level choice. Third, spread of names and sales of goods increase with word-of-mouth and media coverage, and decrease with popularity and market saturation, respectively. However, a key difference is that names benefit from much less media coverage. Although upcoming celebrity names may influence parents' choice, this is a short-lived situation, difficult to unveil in yearly data on name popularity. Hence, we may consider that names benefit from little to no advertising.

**MATHERIALS AND METHODS**
The analogy between the dynamics of names and durable goods invites to a new approach to quantitative modelling of fashion names.

**Macroscopic model of fashion spread**
To model the dynamics of the yearly number of newborns given a certain name (i.e., the yearly incidence of the name), we propose the celebrated Bass model developed for dynamics of sales of durable goods [1]

$$\frac{dN}{dt} = \left[p + q\left(\frac{N}{K}\right)\right](K - N),$$

where *N* is the number of sales, *p* is the *innovation coefficient* (likelihood that someone will start using the product because of media coverage), *q* is the *imitation coefficient* (likelihood that someone will start using the product because of word-of-mouth) and *K* is the market size. Applying the Bass model to names, we neglect media coverage (i.e., *p*=0) and obtain the logistic model [14]. Hence, each fashion name is assigned only two parameters: one summarizing the impact of word-of-mouth on name choices (i.e., *q*, called in this case *maximum growth rate*) and another one representing the maximum number of bearers (i.e., *K*, called in this case *carrying capacity*).

**Data sources**

To test our new modelling framework, we fit French, Dutch and US nationwide databases of yearly name incidence.

*French data.* We obtained two databases on the number of children born in France with each given name, from the French National Institute for Statistics and Economic Studies (INSEE), distributed by Centre Maurice Halbwachs. The first database is "Fichier des prénoms - Edition 1999", provides name time series for the period 1900-1999. The second database is "Fichier des prénoms - Edition 2009" and covers the period 1900-2008. In the compilation of the second database, more names are considered rare and excluded for privacy reasons.

*Dutch data.* We obtained a subset of the "Nederlandse Voornamenbank" (Dutch forenames database) that covers births in the Netherlands over the 1880-2010 period. This subset contains 2,776 names given more than 500 times during the covered period.

*U.S. data.* We downloaded the yearly lists of baby names with at least 5 occurrences for the period 1880-2011, available on the official website of the US Social Security Administration (http://www.socialsecurity.gov/OACT/babynames/limits.html). Using these data, we constructed time series for each name and kept the ones spanning at least 60 years, to maintain means for comparison with the other databases.

**RESULTS**

We performed nonlinear least-square fitting of time series from French, Dutch and American databases with the logistic model. For evaluating goodness of fit, we used the coefficient of determination $R^2$, which, in the case of nonlinear regression, is also known as the Nash-Sutcliff efficiency index [15]. We excluded <30% of the time series from each database based on the following criteria: (1) the least-square fitting algorithm failed to converge for the respective time series; (2) position of the predicted incidence peak was outside the period covered by the time series; and (3) the width of the predicted peak was broader than the period covered by the time series. Remarkably, the goodness of fit for the time series we kept grouped around high values; see the first row of panels in Figure 1. The average fraction of the yearly incidence associated with these data was more than 85% for each database; see Table 1.

We further focused on a subset of names, which we considered to obey well the logistic model. These names were imposed two additional criteria: (a) the $R^2$ of the fit of the whole time series is larger than 0.6, and (b) the $R^2$ of the fit of the time series centred around the incidence peak and truncated to the width of the peak is larger than 0.6. The last criterion insured that goodness of fit did not result only from fitting the wings, but also the peak of the incidence time series. In total, thousands of names qualified, representing up to a third of the total number of the names in each database. Furthermore, the average yearly incidence associated with these names represented more that 50% of the total yearly incidence reported in database; see Table 1.

Inspection of the fitting results yielded two major conclusions. First, plotting goodness of fit versus cumulative incidence revealed that name fashion occurs on many popularity scales; see the second row of panels in Figure 1. That is, high goodness of fit is observed over several orders of magnitude of the cumulative incidence of names. Second, a variety of fashion profiles are possible, as the values for the carrying capacity and imitation coefficient emerging from the fits do not appear correlated; see the third row of panels in Figure 1. While carrying capacity of names varies over 3-4 orders of magnitude, the imitation coefficient remains confined between 0.04 and 1.

Examples of fit for names with different popularities are presented in Figure 1 (black and red correspond to high and low popularity, respectively): the French names Philippe and Francisco in panel G, the Dutch names Ingrid and Moniek in panel H, and the American names Diane and Seymour in panel I. As with the sales of durable goods, the Bass (i.e., reduced to logistic, in our case) model has predictive value for fashion name dynamics. We demonstrate this in Figure 2 for the French name Florine. We used data up to the year 1999 to obtain the fit parameters and their corresponding confidence intervals. Then, we made predictions which were well validated by the 2000-2008 data.

Our model is based on the principles of population-level diffusion dynamics, implicitly incorporating distinct traits of individual-level behaviour. In marketing economics, diffusion dynamics of adopting innovative durable goods are further explained using game models of individual-level choice based on utility functions [2, 3]. However, a similar approach may not hold for the diffusion of first names, as it may be argued that fashion does not have utility beyond inter-human interactions. In the appendix S1, we propose a new paradigm: to describe individual-level choice of fashion names, not subject to media coverage, by combining a game of social interactions with a model of decision making. In the end, this results in the expected logistic dynamic for the name incidence.

**DISCUSSION**
Two strands of sociological research explain how names become available for going into fashion. First, class structure is used to explain how some names, chosen by the upper class to distinguish itself from the lower classes, go into fashion as middle-class parents adopt them, emulating upper-class behaviour. Hence, these names lose distinction and are abandoned by upper-class parents, who search for other names [16]. A second strand of research downplays class difference, postulating that names stem through internal processes, by which a traditional name (e.g., Elizabeth) gives birth to multiple derivatives (e.g., Liza, Liz, Lizzy, Eliza...) [5].

The dynamics of first name incidence has been described using mathematical models. Model parameters should relate naturally to the name dynamics and be few in number, such that a small amount of data is sufficient to estimate them and make model-based predictions possible. However, these two requirements have not been well met in mathematical models of name incidence, so far. On one hand, empirical models fit data well using only a few parameters per time series (e.g., gamma curve fit, using three parameters [12]), but the parameter interpretation remains rather obscure. On the other hand, mechanistic models describe explicitly the process of name choice, but, in doing so, they may rapidly accumulate parameters (e.g., memory and delay model, using seven parameters [7]). Based on an analogy between the spread of innovation and fashion, we proposed the Bass model to describe the dynamics of first name incidence, using the mechanistic theory of diffusion processes previously developed for marketing economics [1-3]. In the case of name dynamics, the Bass model reduces to the logistic model, where each name is determined by only two parameters with transparent mechanistic interpretation.

We showed that the logistic model is very suited for describing the propagation of first name fashion. We identified thousands of names well fit by the model and showed that fashion occurs at all popularity scales. It is well accepted that popular names may be driven by fashion; however, we demonstrated that less popular names could undergo similar dynamics. Whether the same sociology applies equally to small- and large-scale fashion remains a question for the future.

Using our new quantitative approach to modelling population-level spread of first names, we offer a predictive framework. Depending on the size of the time series of a fashion name, it is possible to predict the incidence of that name, answering the question of how popular the fashion name will become and when it will go out of fashion. Furthermore, our modelling approach is consistent with previous studies concerning with the speed of adoption/abandonment of first name fashion; see the appendix S1 for details.

The Bass model is limited to the study of diffusion processes and does not apply to names that are transmitted (i.e., not chosen) from parents to children. However, the model could be further improved to work with other objects of fashion (e.g., roots or endings of first names), and complex social models of communities co-evolving in space and time.

In closing, our work is devoted to how fashion propagates, providing insight in the interdisciplinary problem of the dynamics of choice, a topic of keen interest across sciences.

**SUPPORTING INFORMATION**
**Appendix S1**

**ACKNOWLEDGEMENTS**

We thank INSEE and Centre Maurice Halbwachs for the French name databases: "Fichier des prénoms - Edition 1999" and "Fichier des prénoms - Edition 2009". We also thank Gerrit Bloothooft for the Dutch database.

**Conflict of interest:** None.

**Author Contributions** B.C., V.S. and R.B. conceived the study and contributed to obtaining the data. R.B. analysed the data. All authors contributed conceptually, and wrote, edited or commented on the text. All authors approved the final version of the manuscript.

**Tables**

**Table 1.** Summary statistics of fit results. First, we kept time series having the peak within and not broader than the time period covered by the database; all the others were discarded. Second, we considered time series being well fit by the logistic model, if the $R^2$ of the fit of the whole time series and that of the truncated time series centred around the peak and covering the width of the peak were both >0.6.

| **Database** | **French** | **Dutch** | **American** |
|---|---|---|---|
| Fraction of time series kept | 71% (9157/12917) | 89% (2471/2776) | 72% (3619/4997) |
| Average fraction of the yearly incidence corresponding to the time series kept (min-max) | 85% (61-95%) | 96% (68-99%) | 86% (60-97%) |
| Fraction of well fit time series | 9% (1193/12917) | 31% (872/2776) | 28% (1410/4997) |
| Average yearly proportion of well fit time series (min-max) | 16% (12-22%) | 29% (22-34%) | 32% (27-42%) |
| Average fraction of total yearly incidence corresponding to the well fit time series (min-max) | 75% (45-88%) | 66% (44-73%) | 53% (34-69%) |

**Figures Legends**

**Figure 1.** Results from fitting the logistic model to (the kept) time series of name incidence from French (left column), Dutch (middle column) and American (right column) databases. **First row:** Histograms of goodness of fit $R^2$. **Second row:** Goodness of fit $R^2$ versus cumulative incidence of names. Names with high cumulative incidence tend to be well fit. With decreasing cumulative incidence, goodness of fit spreads over a larger range yet remains high for many names. **Third row:** Parameter sets corresponding to the well fit time series plotted as the carrying capacity, $K$, versus the imitation coefficient, $q$. The two parameters do not appear to be strongly correlated. While $K$ varies over several orders of magnitude, $q$ remains tightly confined between 0.04 and 1. **Fourth row:** Examples of time series of name incidence fit with the logistic model. For each database included in our study, we chose two names with different popularity (high and low popularity names are shown in black and red, respectively). Data is represented with dots and the fit models with lines. The panels are as follows: **J** French names Philippe (black) and Francisco (red); **K** Dutch names Ingrid (black) and Moniek (red); **L** American names Diane (black) and Seymour (red).

**Figure 2.** The logistic model has predictive power. We illustrate the case of the French name Florine. The data up to year 1999 (black dots) were used to obtain the fit parameters; their 95% confidence intervals were estimated through wild bootstrap. The black line represents the best fit and the grey region the uncertainty generated by bootstrap. The data on the period 2000-2008, shown as blue stars, are well predicted by the logistic fit of the data on the previous period.

**Figure 1.**

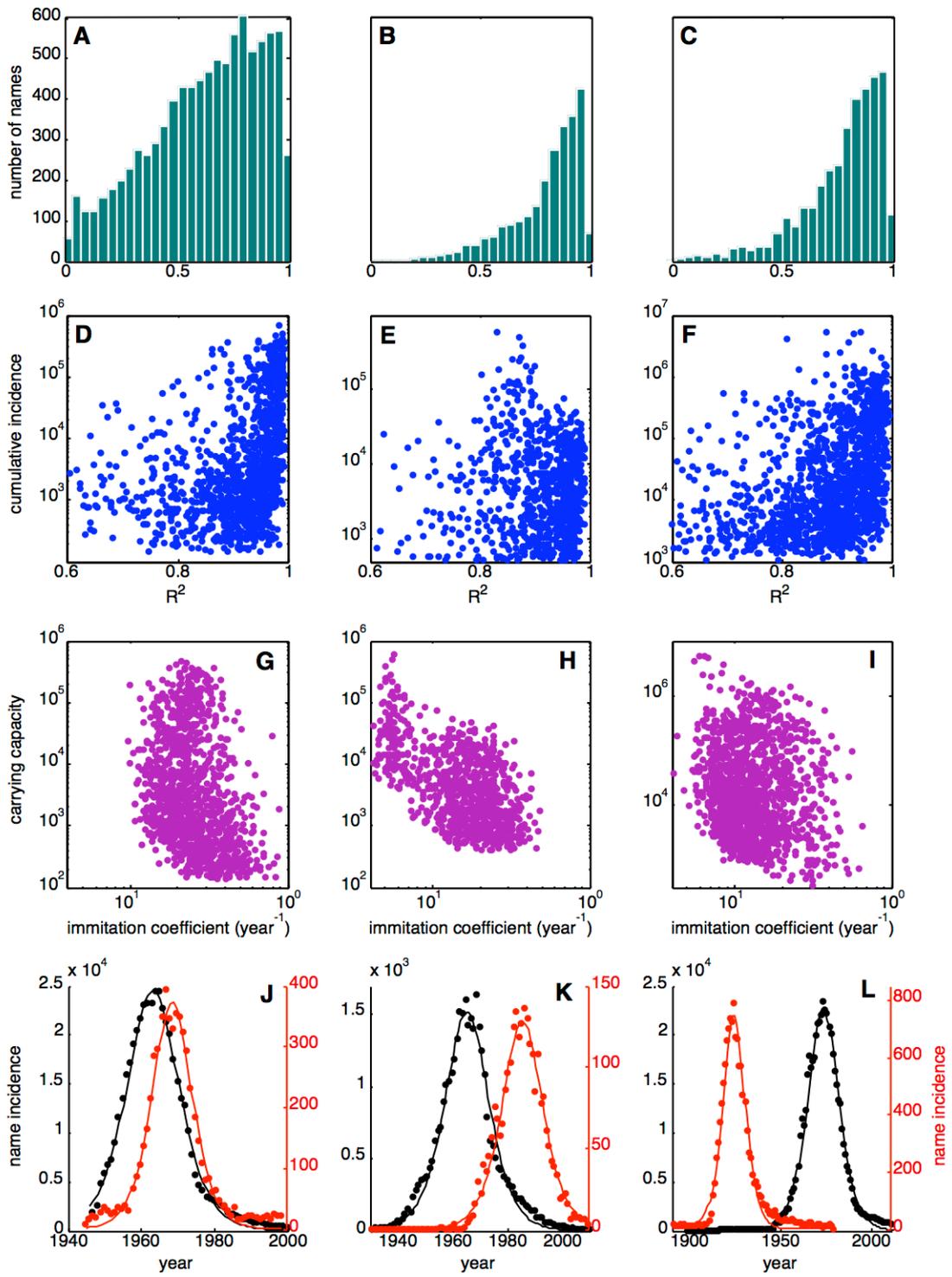

**Figure 2.**

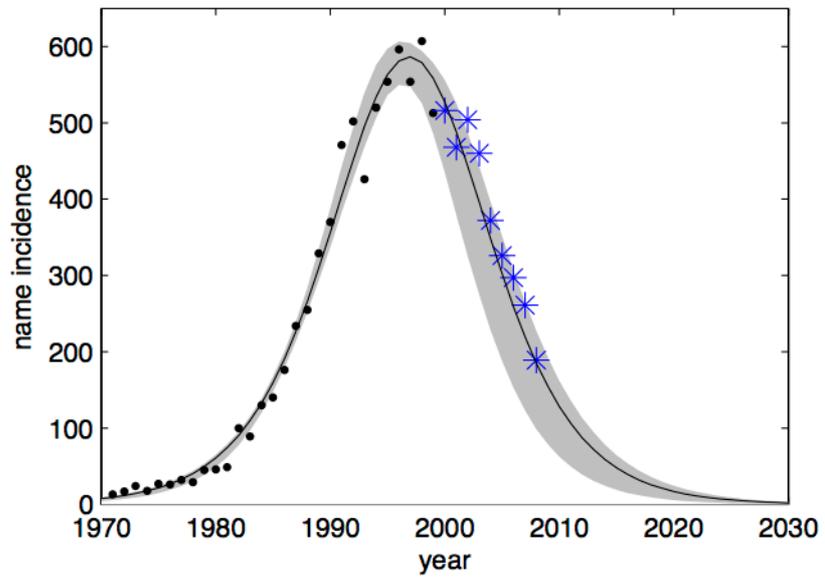

# The diffusion dynamics of choice: From durable goods markets to fashion first names
### SUPPORTING INFORMATION, APPENDIX S1

Baptiste Coulmont,[1] Virginie Supervie[2,3] and Romulus Breban[4]*

[1]*Univ Paris 08, CNRS UMR 7217, Paris F75017, France;*
[2]*Sorbonne Universités, UPMC Univ Paris 06, UMR_S 1136 Pierre Louis Institute of Epidemiology and Public Health, F-75013, Paris, France;*
[3]*INSERM, UMR_S 1136 Pierre Louis Institute of Epidemiology and Public Health, F-75013, Paris, France;*
[4]*Institut Pasteur, UEME, Paris, 75724, France.*
*Romulus.Breban@pasteur.fr

### Paradigm microscopic model of fashion spread

In marketing economics, diffusion dynamics of adopting innovative durable goods are explained using game models of individual-level choice based on utility functions [1, 2]. However, a similar approach may not hold for the diffusion of first names, as it may be argued that fashion does not have utility beyond inter-human interactions. Hence, we propose to describe individual-level choice of fashion names, not subject to media coverage, by combining a game of social interactions with a model of decision-making.

### (a) Model of social interactions
The game provides the pay-off of social interactions of individuals following two strategies: either (a) having chosen the fashion name, or (b) having chosen any other name or being undecided. Of course, individuals do not switch freely between the two strategies at every encounter. At most, undecided individuals may switch once by choosing the fashion name. Hence, the Nash theory of equilibrium does not apply in this case.

The pay-off matrix of the game is given in Table S1 and expresses the role of fashion in social interactions. We assume that when two individuals who did not choose the fashion name meet, the result is a neutral for both (this situation serves as reference). When an individual who did not choose the fashion name meets one who has chosen the fashion name, the pay-off is negative $\beta$ for the first and positive $\alpha$ for second. When two individuals who have chosen the fashion name meet, the pay-off is negative $\gamma$ for each. Under the assumption that the fashion-savvy individuals lose more than the fashion-neutral individuals when encountering someone who has chosen a fashion name (i.e. $\gamma > \beta$), this coordination game is known as the game of *chicken* in economics and political science [3, 4], and *hawks and doves* in evolutionary biology [5].

**Table S1.** Pay-off matrix of the fashion game of social interactions. The parameters $\alpha$, $\beta$ and $\gamma$ are strictly positive and characterize the fashion name.

|  | Fashion name | All other names or undecided |
|---|---|---|
| Fashion name | $-\gamma, -\gamma$ | $\alpha, -\beta$ |
| All other names or undecided | $-\beta, \alpha$ | 0, 0 |

**(b) Individual-level model of choice**

For simplicity, we consider a homogeneous society with constant population size $P$, where the birth rate matches the death rate, and neglect death of the fashion name recipients before the fashion wave has passed. We assume that each individual makes one name choice per lifetime (i.e., two name choices per couple). For this, the individual communicates at time $t$ with others and cumulates advice pro and contra adopting the fashion name. Denote by $N(t)$ the number of individuals who have chosen the fashion name up to time $t$, the total pay-off of the fashion savvy individuals is

$$\Pi_{\text{fashion}}(t) = N(t)\{\alpha[P - N(t)] - \gamma N(t)\},$$

while the pay-off of the fashion-averted individuals is

$$\Pi_{\text{non-fashion}}(t) = [P - N(t)][-\beta N(t)],$$

as quantifiers of the global opinion, pro and contra, respectively, about adopting fashion. The individual decides to adopt a fashion name with a probability proportional to $[\Pi_{\text{fashion}}(t) - \Pi_{\text{non-fashion}}(t)]/P$, which represents the relative advantage of choosing a fashion name over a non-fashion name per consulted individual. Hence, we obtain the following dynamical model for the fashion name popularity

$$\frac{dN}{dt} \propto \frac{1}{P}\left[\Pi_{\text{fashion}}(t) - \Pi_{\text{non-fashion}}(t)\right] = (\alpha + \beta)N\left(1 - \frac{N}{K}\right).$$

This represents the well-known logistic model [6], where $N(t)$ may be reinterpreted as the number of individuals bearing the fashion name at time $t$. The parameter $K$ is known as the carrying capacity of the name and is given by

$$K = \frac{P(\alpha + \beta)}{\alpha + \beta + \gamma}.$$

An important limitation of our individual-level model of decision-making is that it focuses on one particular fashion name only, ignoring the competition that exists between thousands of fashion names at one moment in time. Thus, the above modeling exercise remains just a proof of concept. A more detailed model is a topic for future work.

# Relation between adoption and abandonment of names described by the logistic model

The speed of adoption/abandonment of a positive signal $I(t)$ is defined using the *local exponential growth rate* [7], which is a function denoted by $k(t)$ that helps representing $I(t)$ in the following form

$$I(t) = I(0)e^{k(t)t}.$$

Writing $k(t)$ as

$$k(t) = \frac{1}{p}\ln[1+r(t)/100],$$

introduces the function $r(t)$, which is called the percent increase rate of $I(t)$ over the period $p$; $r(t)$ is also called *speed of adoption (abandonment)* if $r(t)$ is positive (negative) [7].

We now calculate $r(t)$ for the incidence signal of the logistic model

$$I(t) \equiv \frac{dN(t)}{dt} = qN(t)\left[1 - \frac{N(t)}{K}\right],$$

where we choose the logistic function in the form

$$\frac{N(t)}{K} = \frac{1}{1+e^{-qt}},$$

such that the maximum incidence is attained at $t=0$. Straightforward algebra yields

$$\frac{r(t)}{100} = \left[\frac{4e^{-qt}}{(1+e^{-qt})^2}\right]^{p/t} - 1.$$

We now compute the ratio between the speed of adoption at time $\tau$ *before* the incidence peak and the speed of abandonment at time $\tau$ *after* the incidence peak as a measure of their interdependence; see Figure S1.

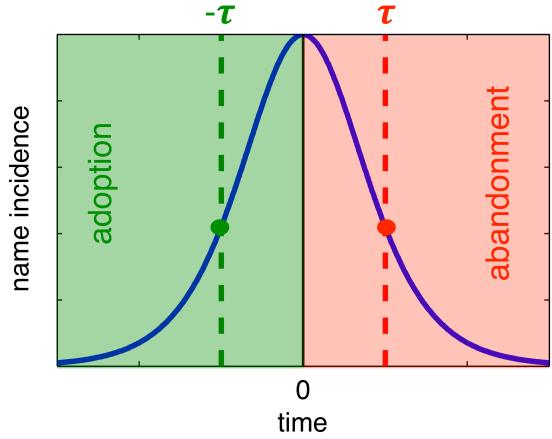

**Figure S1.** Regimes of adoption and abandonment of a name whose incidence dynamics follow the logistic model.

We obtain

$$\eta(\tau;q) \equiv \frac{r(-\tau)}{-r(\tau)} = \left[\cosh\left(\frac{q\tau}{2}\right)\right]^{2p/\tau}.$$

As $\tau$ takes values from 0 to infinity, $\eta(\tau;q)$ is monotonically increasing from 1 to $e^{qp}$. The time series studied in the main text consisted of yearly data (i.e., $p=1$) and, if well fit by the logistic model, had $q$ confined in the interval (0.04, 1). Hence, in this case, the speed of adoption at time $\tau$ before the incidence peak and that of abandonment at time $\tau$ after the incidence peak are closely related, as their ratio lies in a fairly narrow interval; i.e., from $(1, e^{qp}) \approx (1, 1.04)$, if $q=0.04$, to $(1, e^{qp}) \approx (1, 2.72)$, if $q=1$.